# Alloyed β-(Al$_x$Ga$_{1-x}$)$_2$O$_3$ Bulk Czochralski Single β-(Al$_{0.1}$Ga$_{0.9}$)$_2$O$_3$ and Polycrystals β-(Al$_{0.33}$Ga$_{0.66}$)$_2$O$_3$, β-(Al$_{0.5}$Ga$_{0.5}$)$_2$O$_3$) and Property Trends

Jani Jesenovec[1-2], Benjamin Dutton[1-2], Nicholas Stone-Weiss[1], Adrian Chmielewski[3], Muad Saleh[1-2], Carl Peterson[4], Nasim Alem[3], Sriram Krishnamoorthy[4], John S. McCloy[1-2*]

[1]*Institute of Materials Research, Washington State University, Pullman WA, USA 99164-2711*
[2]*Materials Science & Engineering Program, Washington State University, Pullman, WA, 99164, USA*
[3]*Department of Materials Science and Engineering, Materials Research Institute, The Pennsylvania State University, University Park, USA*
[4]*Materials Department, University of California, Santa Barbara, CA 93106, USA*

*Corresponding author:* john.mccloy@wsu.edu

ORCID:
Jesenovec: 0000-0002-5937-6657
Dutton: 0000-0003-1272-130X
Stone-Weiss: 0000-0001-7139-1940
Chmielewski: 0000-0003-2373-5061
Saleh: 0000-0002-1391-2784
Krishnamoorthy: 0000-0002-4682-1002
McCloy: 0000-0001-7476-7771

# Abstract

In this work, bulk Czochralski-grown single crystals of 10 mol.% $Al_2O_3$ alloyed $\beta$-$Ga_2O_3$ – monoclinic 10% AGO or $\beta$-$(Al_{0.1}Ga_{0.9})_2O_3$ – are obtained, which show +0.20 eV increase in the bandgap compared to unintentionally doped $\beta$-$Ga_2O_3$. Further, growths of 33% AGO – $\beta$-$(Al_{0.33}Ga_{0.67})_2O_3$ – and 50% AGO – $\beta$-$(Al_{0.5}Ga_{0.5})_2O_3$ or $\beta$-$AlGaO_3$ – produce polycrystalline single-phase monoclinic material ($\beta$-AGO). All three compositions are investigated by X-ray diffraction (XRD), Raman spectroscopy, optical absorption, and $^{27}Al$ nuclear magnetic resonance (NMR). By investigating single phase $\beta$-AGO over a large range of $Al_2O_3$ concentration (10-50 mol.%), broad trends in lattice parameter, vibrational modes, optical band gap, and crystallographic site preference are determined. All lattice parameters show a linear trend with Al incorporation. According to NMR, aluminum incorporates on both crystallographic sites of $\beta$-$Ga_2O_3$, with a slight preference for the octahedral ($Ga_{II}$) site, which becomes more disordered with increasing Al. Single crystals of 10% AGO were also characterized by X-ray rocking curve, transmission electron microscopy, purity (glow discharge mass spectroscopy, X-ray fluorescence), optical transmission (200 nm – 20 μm wavelengths), and resistivity. These measurements suggest that electrical compensation by impurity acceptor doping is not the likely explanation for high resistivity, but rather the shift of a hydrogen level from a shallow donor to a deep acceptor due to Al alloying. Bulk crystals of $\beta$-$(Al_{0.1}Ga_{0.9})_2O_3$ have the potential to be ultra-wide bandgap substrates for thin film growth, with a lattice parameter that may even allow higher Al concentration $\beta$-$Ga_2O_3$ single crystal thin films to be grown.

# I. Introduction

Ultra-wide bandgap β-(Al$_x$Ga$_{1-x}$)$_2$O$_3$ (AGO) alloyed bulk substrates are needed to improve lattice matching for thin film growth of high alumina β-Ga$_2$O$_3$ by techniques such as molecular beam epitaxy (MBE) and metalorganic chemical vapor deposition (MOCVD). Rapid development of β-Ga$_2$O$_3$ materials and devices is possible due to readily available bulk material grown by techniques such as Czochralski, vertical gradient freeze (VGF), edge-defined film-fed growth (EFG), vertical Bridgman, and float zone techniques.[1–4] The monoclinic phase of alumina, θ-Al$_2$O$_3$, is predicted to have a bandgap of 7.24 eV,[5,6] compared to β-Ga$_2$O$_3$ (4.5 – 4.9 eV).[7] The larger bandgap of AGO materials would enable higher critical field devices[8,9] and electronics for deeper in the ultraviolet.[10,11] Further, *n*-type doping (e.g., Si, Zr, Hf) is still possible in AGO material,[12] and has been reported for thin films.[13,14] Modulation-doped two-dimensional electron gas AGO/Ga$_2$O$_3$ devices with silicon delta doping are critical for improving mobility and scaling up for high performance transistors.[15–22]

Epitaxial growth of β-(Al$_x$Ga$_{1-x}$)$_2$O$_3$ (AGO) has been explored using MBE,[23] and MOCVD.[14,22] Kaun *et al*. reported that the monoclinic phase stability limit of Al$_2$O$_3$ in β-Ga$_2$O$_3$ (010) at 600 °C growth by MBE was less than ~18%.[24] Higher aluminum incorporation is reported in MOCVD growth, with the maximum aluminum incorporation dependent on the growth orientation. Higher Al incorporation was observed in AGO films grown on (100)[25] and (-201)[26] substrates, compared to (010) orientation,[14,22] with a maximum Al of 52% on the metal site.

Bulk CZ grown crystals of AGO have been synthesized previously at doping concentrations up to 5 mol.% Al$_2$O$_3$ β-(Al$_{0.05}$Ga$_{0.95}$)$_2$O$_3$ with demonstrated 0.11 eV increase in the bandgap from β-Ga$_2$O$_3$.[27] Additions of low concentration of Al$_2$O$_3$ had a negligible effect on melting temperature, only 16 K higher than unintentionally doped (UID) β-Ga$_2$O$_3$.[27] Alloying with alumina also reduced the decomposition of the melt at growth temperatures.[27] AGO has also been studied during co-doping experiments with Ce + Si in order to increase the bandgap and improve scintillation by tailoring *n*-type conduction.[28] The optical floating zone method has been used to grow AGO up to β-(Al$_{0.15}$Ga$_{0.85}$)$_2$O$_3$, with the purpose of increased

X-ray scintillation through increasing the bandgap.[29] Grown by Li *et al.*, these crystals achieve good detector performance, which is attributed to their materials' insulating properties and high quality of the grown crystal. Recent work on CZ grown crystals has increased the Al concentration further, up to β-$(Al_{0.182}Ga_{0.818})_2O_3$ (described in Ref[30] as 7.5 at% Al by energy dispersive spectroscopy (EDS), but identical to $x = 0.182$ as shown in Ref[31]) and most recently β-$(Al_{0.23}Ga_{0.77})_2O_3$ (9.2 at% Al by EDS),[31] though CZ crystals with even $x = 0.062$ show evidence of polycrystallinity from X-ray rocking curve measurements.[31] Crystals of $Ga_2O_3$ alloyed with higher $Al_2O_3$ fraction grown by Czochralski have not been reported to our knowledge.

In the current work, we describe the growth attempts of AGO with $Al_2O_3$ contents from 10-50 mol.%. The intent of the described study is to report the trends in structural and spectroscopic properties of high-$Al_2O_3$ AGO, which may be polycrystalline, as compared to lower-$Al_2O_3$ AGO which shows bulk single crystallinity. Here we report successful synthesis of β-$(Al_xGa_{1-x})_2O_3$ bulk single crystals by the Czochralski technique, at $x = 0.1$ (10% $Al_2O_3$, hereafter referred to as 10% Al AGO), and further studies on growths where bulk single crystals were not obtained, for compositions where $x = 0.5$ (50% $Al_2O_3$) and 0.33 (33% $Al_2O_3$). 10% Al on the metal sites of β-$Ga_2O_3$ yielded bulk single crystals of sizes sufficient to act as substrates for thin film growth while increasing the bandgap appreciably, and demonstrated several orders of magnitude higher resistivity than UID material. Comparison of the three compositions allowed insight into structural changes with increasing $Al_2O_3$ addition, including the dependency of the lattice parameters of the monoclinic phase on $Al_2O_3$ concentration. Nuclear magnetic resonance (NMR) indicated Al incorporation on both the octahedral ($Ga_{II}$) and tetrahedral ($Ga_I$) sites, with a preference for the $Ga_{II}$ site for all compositions. XRD and Raman measurements show clear trends associated with Al incorporation into β-$Ga_2O_3$.

## II. Methods

Growths of β-$(Al_xGa_{1-x})_2O_3$, where $x$ = 0.5, 0.33, 0.1, were performed from the melt with methods similar to those previously published.[32–34] High purity precursor powders – 5N (99.999%) $Ga_2O_3$ (ABSCO Limited, Haverhill, Suffolk, UK) and 4N7 (99.997%) $Al_2O_3$ (Inframat Adv. Mat., Manchester, Connecticut, USA) – were batched and mixed for 18 hours at 50 rpm in a rotary mill. Two charges of ≈425 g were prepared, cold pressed at 140 MPa, and calcined in an alumina crucible with Pt foil lining at 1600°C or 1500°C in air for 15 h. In preparation of the 50 mol.% $Al_2O_3$ and 33 mol.% $Al_2O_3$ growths, 1500°C calcines were used, which did not sufficiently sinter the charge due to the addition of alumina, thus for 10 mol.% $Al_2O_3$ a 1600°C calcine was conducted which sintered the charge properly. The first charge was melted and cooled with material loss of ≈17 g due to evaporation, and then the second charge was added with a total AGO material weight of ≈833 g. The growths were conducted in an iridium crucible (Johnson Matthey, London, UK) with a 70 mm diameter, 70 mm height, 2 mm wall thickness and 3 mm bottom thickness, inductively coupled to a radio frequency (RF) coil. The crucible was rotated at 2 rpm for the duration of the run, with no rotation of the seed. Standard oxygen flow scheme for $Ga_2O_3$ was employed to reduce decomposition of the melt, following the work by Galazka et al.,[1,35] in which a mixed gas of Ar+10 vol.% $O_2$ was used when the melt was at growth temperatures; otherwise when > 1100°C, the gas environment was Ar+2.5-3.5 vol.% $O_2$, and at < 1100°C, $O_2$ < 0.2 vol.% was used. During growth, a 10 – 20 kPa overpressure was maintained. Temperatures were monitored during growth using two pyrometers, an Ircon 2 channel pyrometer, and a Sekidenko OR1000F pyrometer, and these measurements demonstrated good agreement in temperature throughout the run. Precise temperature measurement of the growths has been described elsewhere.[33]

At ≈1850 °C, crystals were grown by CZ at 2 mm/h pull rate and 2 rpm rotation, and subsequently by VGF with no rotation upon cooling at 1-2 °C $min^{-1}$. Non-spiral 10% Al AGO CZ crystals were grown on a Mg-doped β-$Ga_2O_3$ seed, with the boule exhibiting a diameter of 41.5 – 36.6 mm and a cylindrical height of 14.5 mm. The height from tip to heel was 22.4 mm. 50% and 33% Al AGO growths did not yield successful CZ crystals, with the pulled CZ mass and most of the VGF being highly polycrystalline.

For single crystalline samples of 10% Al AGO, (100)-oriented crystals of size of 0.25 × 0.25 cm$^2$ to 1 × 1 cm$^2$ with varying thicknesses were obtained from the VGF and CZ growths. 33% AGO had very poor single crystalline quality, yielding only a few samples large enough to be tested, while the rest were polycrystalline. Polycrystalline material was obtained from the VGF boule of the 50% Al and 33% Al AGO and ground up for measurements. The material was single phase for all growths as confirmed by XRD (see Results). Polycrystalline growths were opaque and white in color. 10% Al VGF crystals were greyer than comparable UID β-Ga$_2$O$_3$. The 10% Al CZ pulled crystal was beige in color, similar to our UID β-Ga$_2$O$_3$.

High resolution High Angle Annular Dark Field-scanning transmission electron microscopy (HAADF-STEM) imaging and electron diffraction were carried out on the 10% Al AGO using a FEI Titan G2 60-300 transmission electron microscope (TEM) operated at 300 kV. A condenser aperture of 70 μm was used with a corresponding convergence angle of 30 mrads and the inner and outer annular detector collection semi-angles equal to 42 and 244 mrad, respectively. The probe current was approximately 90 pA. Camera length for both imaging and diffraction was set to 115 mm. Diffraction pattern indexing was done with the help of SingleCrystal.

Phase purity and lattice parameters were analyzed for all compositions with an X-ray diffractometer (XRD) using Cu K$_α$ radiation (λ = 1.5406 Å), either Panalytical X'pert Pro at 45 kV and 40 mA (UID, 10% AGO, 33% Al AGO), or a Rigaku Miniflex 600 at 40 kV and 15 mA (50% Al AGO). In all cases, scans were performed from 5-90° 2θ with a step size of 0.5° and 10 s per step. Final XRD patterns were obtained from summation of five sequential scans. Rietveld refinement was conducted in HighScore Plus (Malvern Panalytical) software with custom refinement parameters. High resolution rocking curve measurements were collected on 10% Al AGO with a 4-bounce Ge (220) monochromator and a PiXCEL3D (Malvern Panalytical) X-ray detector using the (400) reflection and 45 kV and 40 mA settings, and 1 mm receiving slit size.

Single resonance $^{27}$Al magic angle spinning nuclear magnetic resonance (MAS NMR) spectra were recorded on a 14.1 T Varian DD2 600 MHz spectrometer using a commercial 4.0 mm MAS NMR probe (Agilent). All compositions were powdered then packed into 4.0 mm zirconia rotors and spun at 15 kHz. The spectra were measured at 156.27 MHz resonance frequencies with π/6-pulse durations of 0.8-1.2 μs and recycle delays of 2-16 seconds. Measurements were signal-averaged over at least 150 scans and were processed without additional line broadening. Chemical shifts of $^{27}$Al were reported relative to powdered $AlPO_4$, measured at 40.7 ppm relative to 1 M $Al(NO_3)_3$ at 0 ppm. The spectra were fitted using CZSimple models in DMFit software to simulate the quadrupolar lineshapes observed and quantify the Al site preference.[36]

Raman shift and peak broadening was studied using a Raman microscope; for all samples, a Thermo Fisher DXR2XI with a 532 nm laser and a grating with a Raman shift range of 100 – 3500 cm$^{-1}$ with a 2 cm$^{-1}$ resolution was used. All samples were (100), although orientation of the 50% Al AGO sample could not be assessed due to the polycrystallinity of the sample. Laser intensity as well as exposure was kept the same between measurements. All measurements are composed of 5 spectra which are then averaged to reduce noise.

Sample purity and aluminum concentration were assessed only on 10% Al AGO, using both glow discharge mass spectroscopy (GDMS) and X-ray fluorescence (XRF) at EAG Laboratory (California, USA). Samples were taken from both the VGF and CZ portions of this growth. GDMS was performed on 10% Al AGO after the crystals were crushed in an alumina mill, while XRF was only performed after crushing in a tungsten carbide mill to verify the incorporation of Al without potential Al contamination. The results summarized in the supplementary material show elements at concentrations above the detection limit, of note an as-expected Al incorporation, and lower than typical incorporation of key acceptors and donors (e.g., in the CZ, Fe: $1\times10^{17}$ cm$^{-3}$, Si: $1\times10^{18}$ cm$^{-3}$).

Room temperature optical transmission measurements in the ultraviolet-visible-near infrared (UV-Vis-NIR) and infrared (IR) were conducted on 160 – 180 μm or 3 mm thick single crystal windows, respectively. Band edge measurements for 10% Al AGO windows were obtained using a Cary 5 UV-Vis-

NIR, for 200 – 3300 nm (50,000 – 3,333 cm$^{-1}$) using a polarizer to obtain transmission for E||b and E||c. IR measurements were obtained using a Bruker Alpha Fourier transformed infrared (FTIR) spectrometer, for 1.3 – 25.0 μm (7,500 – 400 cm$^{-1}$). FTIR spectra were taken before and after a 15 hour 1000 °C anneal in flowing oxygen where samples were placed in an alumina boat with a platinum liner. Optical transmission and Kubelka-Munk scattering measurements, from 190 – 1100 nm (50,000 – 9,090 cm$^{-1}$), were also obtained on powder samples to allow assessment of polycrystalline materials, using an integrating sphere in a UV-Vis spectrometer (ThermoFisher Evolution 260 Bio). All optical spectra were analyzed with respect to absorptions due to the band edge, free electron carriers, impurities and multiphonon processes.

For electrical measurements, 50-50% Ga-In ohmic contacts were placed on both sides of the (100) plane of the 10% Al AGO crystal in a two-point configuration. Samples were annealed at 950°C for 15 minutes, and then more contact material was placed on top of the old contact. This procedure has been shown to produce contacts with ohmic behavior down to temperatures as low as ≈20 K.[34] Two-point through thickness current-voltage (*I-V*) and resistance measurements were obtained with a high impedance picoammeter.

## III. Results and Discussion

### A. Growth

Bulk CZ growth of 10% Al AGO (i.e., β-(Al$_{0.1}$Ga$_{0.9}$)$_2$O$_3$ or β-Al$_{0.2}$Ga$_{1.8}$O$_3$) demonstrated less melt evaporation than UID β-Ga$_2$O$_3$, as expected from previous reports from literature.[27] Supercooling of the crystal by approximately 40°C, from 1850°C to 1810°C, was required in order to seed and grow. Phase diagrams of the β-Ga$_2$O$_3$ and Al$_2$O$_3$ system by Hill *et al.*[37] show a region of reduced solubility below 800°C; however, when cooling the crucible and crystals at a rate of 1°C min$^{-1}$ (specifically at 800°C), 10% Al AGO was realized with no secondary phases.

Obtained crystals are shown in Fig. 1. The CZ boule was asymmetric, although not spiral. Many large grain single crystals separated by grain boundaries and cracks made up the CZ boule, caused by difficulties in seeding. The lack of an after heater in this growth is also considered deleterious to the quality of the crystal, specifically resulting in cracking and a high vertical temperature gradient. Some inclusions are visible on the outside CZ surface, but morphology is similar to Ir deposits typically seen on other CZ $\beta$-$Ga_2O_3$ growths. There are no visible inclusions through the (100) surface of substrates.

10% Al AGO CZ crystals were measured with XRF and indicated an incorporation of 11.74 mol.% $Al_2O_3$, resulting in an effective incorporation coefficient, $mk_{eff}$, of 1.17, agreeing well with literature, where Al is reported having an $mk_{eff}$ of 1.1 in 5 mol.% $Al_2O_3$ in $\beta$-$Ga_2O_3$.[27] There were minor differences in Al concentration between the VGF and CZ boule; however, as shall be shown within this work, the concentrations were such that structural, optical and electronic properties were much the same between the VGF and CZ. Purity data is shown in the supplementary materials.

Polycrystalline growths of 50% Al AGO (i.e., $\beta$-$(Al_{0.5}Ga_{0.5})_2O_3$ or $\beta$-$AlGaO_3$) and 33% Al AGO (i.e., $\beta$-$(Al_{0.33}Ga_{0.67})_2O_3$ or $\beta$-$Al_{0.66}Ga_{1.34}O_3$) were obtained on a sapphire seed and an Ir rod, respectively, as seeding with Mg:$\beta$-$Ga_2O_3$ was unsuccessful due to seed melting. The resultant pulled crystals were polycrystalline and not cylindrical. VGF growth of 50% Al AGO yielded columnar structures throughout the boule, with no single crystals of appreciable size. 33% Al AGO yielded some very small single crystals with poor surface quality consisting of many cracks and uneven surfaces, but otherwise was a mass of polycrystalline material.

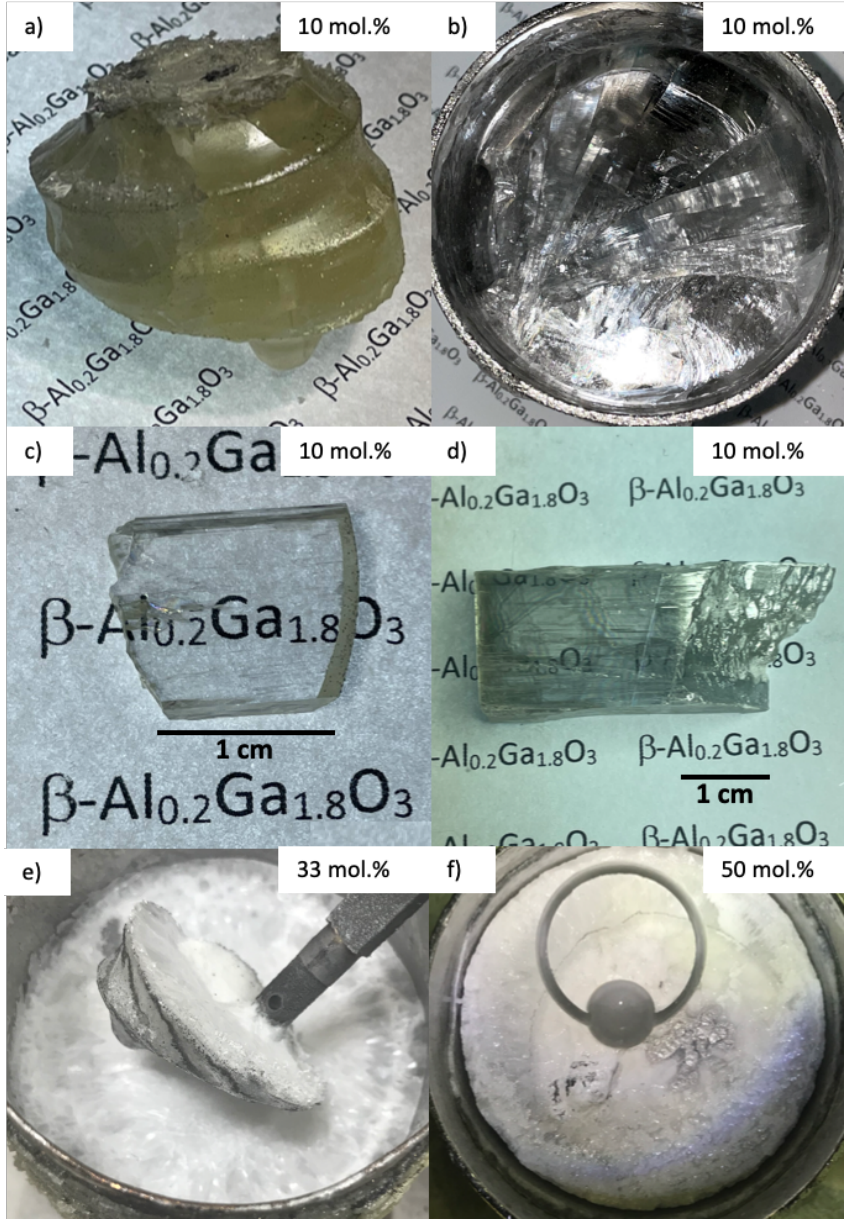

FIG. 1. a) 10% Al AGO CZ pulled crystal. b) 10% Al AGO VGF boule in iridium crucible. c) As-grown chunk of single crystal from the CZ boule. d) As-grown chunk of single crystal from the VGF boule with clearly rough surface. e) 33% Al AGO CZ pulled crystal seeded on Ir rod and VGF boule. f) 50% Al AGO VGF boule with various core drill holes in the surface. Note metallic spot near center of boule, most likely Ir.

### B. Structural Properties

In order to assess crystal quality in terms of lattice defects, second phases, and Al incorporation, several techniques were applied to AGO samples.

HAADF-STEM was applied in order to assess defects and crystal quality of the 10% Al AGO crystals. Images (Fig. 2) of the investigated area indicate a well-ordered lattice with no visible structural defects, and successful Al incorporation onto both octahedral and tetrahedral sites in the expected monoclinic crystal structure.

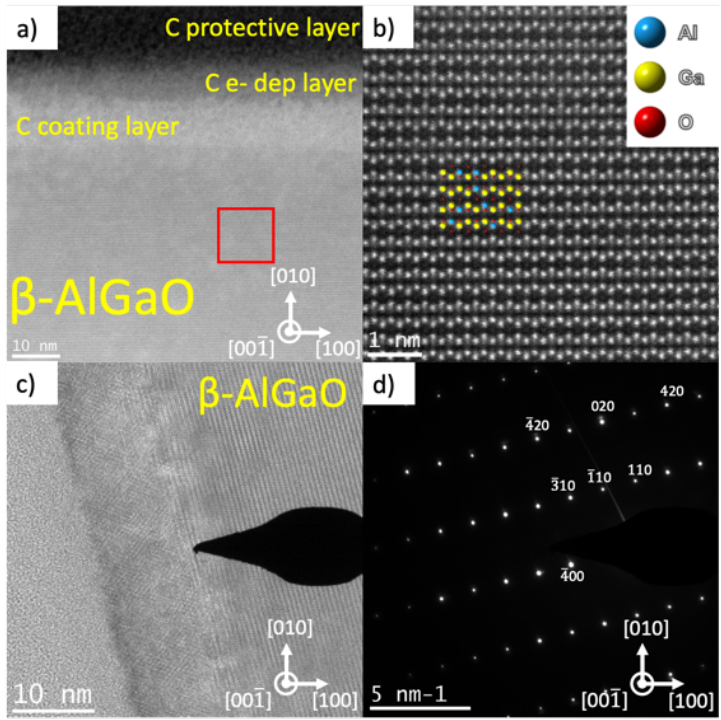

FIG 2. a) High resolution High angle annular dark field-scanning transmission electron microscopy (HAADF-STEM) image of 10% Al AGO in the [00$\bar{1}$] projection. The top of the image corresponds to the different carbon coating and protective layers. b) Zoomed-in image of area highlighted by the red square in Fig 2.a). c) Low magnification TEM image of the sample in the [00$\bar{1}$] projection and its corresponding diffraction pattern is shown in d). These different TEM techniques show the high quality and single crystallinity of the sample.

High resolution XRD rocking curve measurements on as-grown and as-cleaved UID and 10% Al AGO samples demonstrate crystal quality expected from Czochralski grown single crystals (Fig. 3). Only one sample was measured from each boule, although attempts were made to select from the regions representative of the measurements presented herein. Industrially grown β-Ga$_2$O$_3$ EFG substrates, for example, are certified for ≤ 150 arcsec (Novel Crystal Technologies, Inc.). A (010) Fe-doped β-Ga$_2$O$_3$ substrate obtained from industry and grown via CZ was measured with the same XRD system and demonstrated ≈105 arcsec. Research-grade β-Ga$_2$O$_3$ crystals have shown 17 arcsec (EFG),[3] 10 – 50 arcsec (vertical Bridgman),[38] and 23 – 36 arcsec (Czochralski).[35] Previous reports of β-(Al$_{0.0625}$Ga$_{0.9375}$)$_2$O$_3$ via CZ showed lower quality crystals at lower Al concentration with split diffraction peaks, and although higher Al concentration ($x$ = 0.23) was reported, rocking curve was not shown.[31] Of note in Fig. 3 is the sideband right of the primary peak, as well as general asymmetry of the peaks. This is caused by the cleavage planes or cracking in β-Ga$_2$O$_3$ which cause mosaic-like or platelet-like surface morphology at large length scales, or the presence of low angle grains and twin boundaries in the crystal.

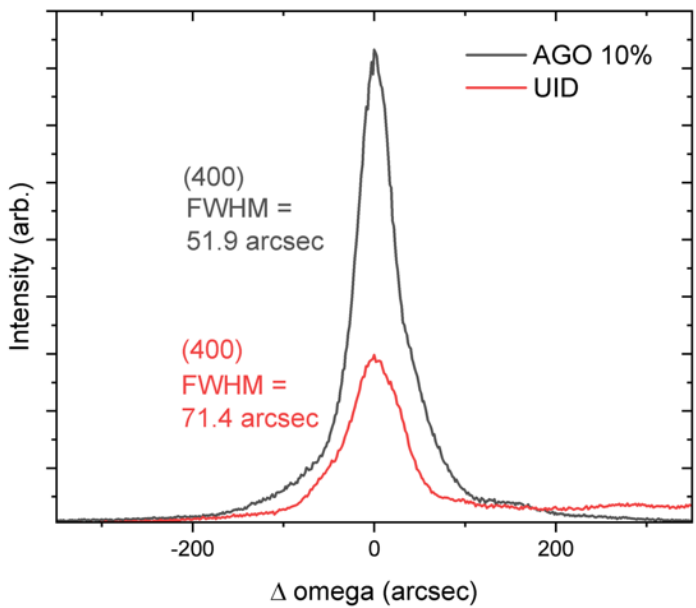

*FIG 3. High resolution rocking curve XRD measurement of (100) oriented AGO 10% and UID. Receiving slit on PiXCEL3D was 1 mm.*

Powder XRD was used to study lattice behavior as a function of aluminum content. As aluminum concentration increases, a shift to higher 2θ angles is noticeable in the patterns, attributed to shrinking of the lattice (Fig. 4a,b). In order to assess whether second phases were forming, XRD was applied to material obtained from the top and bottom of the crucible, and from all AGO concentrations attempted, and no significant differences (other than shifts) between the patterns were found nor matched any suspected phases (Al$_2$O$_3$ or contaminants) in the XRD libraries. Several peaks change in intensity as a function of

AGO concentration: 62.7° 2θ increased and 27.1° 2θ decreased with increasing Al content. These peaks were matched to the expected monoclinic β-$Ga_2O_3$ structure, and are most likely indicative of disorder within the crystal especially at high Al concentration. Measurements on all of the alloys demonstrated patterns with peaks identifiable by the UID β-$Ga_2O_3$ structure, indicating the alloys formed solid solutions with no second phases. A higher resolution plot of Fig. 4b may be found in the supplementary. Due to the availability of high Al concentration single phase polycrystals at $x = 0.33$ and $x = 0.50$, the authors were able to study the crystal lattice as a function of alloying. Lattice parameters were obtained via Rietveld refinement (Fig. 4c), with a linear decrease in lattice size as expected based on 2θ shift. These trends suggest that future studies could use lattice parameters to directly obtain the Al concentration on material with an unknown $x$ using the fitted equations presented here. This would allow for a benchtop approximation of $x$, complementary to other techniques such as energy dispersive spectroscopy (EDS) or X-ray fluorescence (XRF).

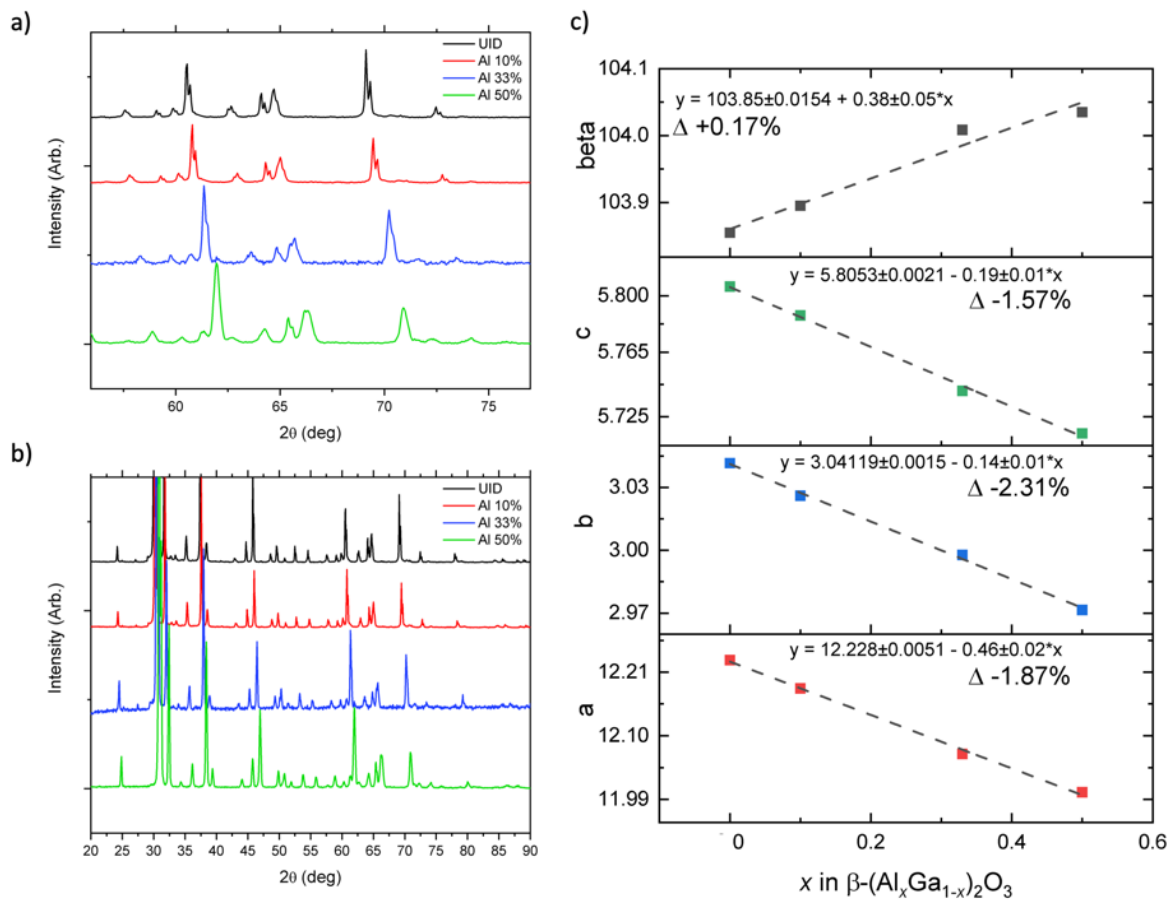

FIG 4. a) XRD patterns of UID vs. AGO, for a region where the peak shift is evident. b) XRD patterns of UID vs. AGO. All spectra in a,b) have been offset for clarity. c) Lattice parameters as a function of $Al_2O_3$ doping. Lines shown are linear fits to the data.

$^{27}$Al MAS NMR spectra were collected to investigate the site preference of Al in β-$Ga_2O_3$ as a function of aluminum content. It is observed in Fig. 5 that each sample contains Al in both tetrahedral (Al-4; centered near 68 ppm) and octahedral (Al-6; centered near 9-12 ppm) sites. Samples were fitted using three distinct peaks within the Al-4 resonance and one to three distinct peaks within the Al-6 resonance (complete fitting parameters provided in Table S1). Area integration of the fitted quadrupolar lineshapes produces site fractions for each sample studied, indicating that AGO with 10% Al (CZ and VGF), 33% Al, or 50% Al all contain a majority of Al in the octahedral ($Ga_{II}$) site, as is similar to the observations by Cook et al.,[39] with the area fractions 64%, 71%, and 62%, respectively. Further, we observe insignificant differences in Al-6 fraction between the AGO 10% Al CZ and VGF samples (64% vs. 63% Al-6),

considering that ±1% error is typically assessed to quantifying site fractions in MAS NMR. Notably, the Al-6 resonance of the NMR spectra of 33% Al AGO and 50% Al AGO appear significantly broadened and shifted upfield as compared to the 10% Al AGO samples, and the former samples were satisfactorily fitted using one broad peak, while the latter had to be fit using three distinct peaks. The broadening of the Al-6 peak is attributed to a more disordered octahedral Al site in the higher Al-doped samples; this is confirmed by the increasing value of the quadrupolar coupling constant with increasing Al (see Table S1). The results show that Al prefers to enter the $Ga_{II}$ (octahedral) site but also enters the $Ga_I$ (tetrahedral) site to a significant extent, and that with increased Al concentration, the average octahedral site containing Al becomes more disordered. These results are in agreement with work conducted on thin films, showing similar incorporation behavior.[40]

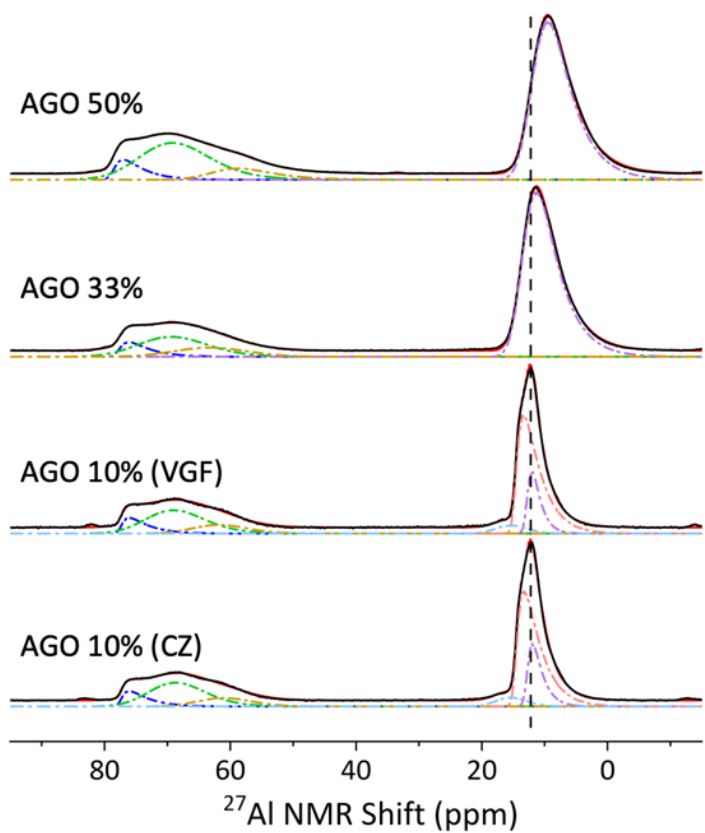

FIG 5. $^{27}$Al MAS NMR spectra of AGO samples. Spectra were fitted using CZSimple functions in DMFit software to fit the quadrupolar lineshapes.

Raman spectroscopy also evidenced a decrease in the lattice volume as scattering shifted to higher wavenumbers with increasing Al concentration (Fig. 6). The Raman scattering shift from UID to 50% Al AGO was 35-40 cm$^{-1}$. Broadening of Raman peaks indicated more structural disorder as Al content increased. Assignment of Raman modes was performed based on Onuma *et al.*[41] Severe broadening occurred at 50% Al AGO, where several peaks merged ($B_g(5)$, $A_g(8)$, $A_g(9)$: 600 – 700 cm$^{-1}$), became shoulders ($B_g(2)$, $A_g(2)$, $A_g(3)$: 150 – 200 cm$^{-1}$), or broadened into the background ($B_g(4)$, $A_g(6)$, $A_g(7)$: 400 – 500 cm$^{-1}$). Bulk second phases would appear as new peaks or features, which were not present in data taken over several representative samples. Specifically, these spectra are distinct from $Al_2O_3$ Raman spectra, for example found in the RUFF Raman library[42] and thus are indicative of AGO remaining in the monoclinic $\beta$-$Ga_2O_3$ structure rather than a change to the corundum, which was also verified by XRD.

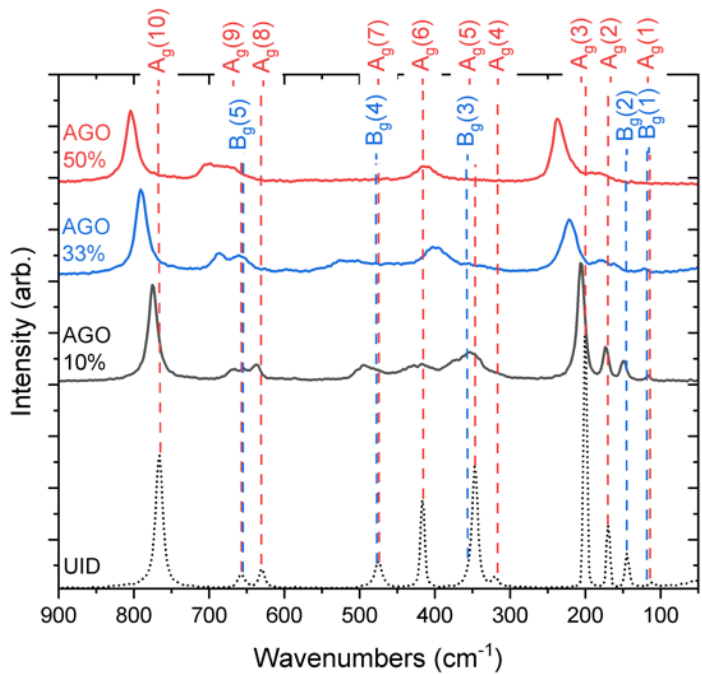

*FIG 6. 532 nm unpolarized Raman spectra taken with the same laser power and exposure with Raman modes assigned, indicating an increase in disorder and shrinking of the lattice as a function of Al concentration. Plots were offset in the y-axis only for ease of viewing.*

## C. Optical Properties

Aluminum concentration caused an increase in the apparent bandgap of β-$Ga_2O_3$ (Fig. 8), as also shown in literature.[5,27–29,43] Free carrier absorption was notably absent in aluminum alloyed material compared to UID β-$Ga_2O_3$ grown by the same method, which shows free carrier absorption from unintentional shallow donor impurities (Fig. 7), in agreement with previous studies on Al doping in β-$Ga_2O_3$.[29] This lack of free carrier absorption indicates that the samples are insulating, with near infrared spectra similar to those seen in acceptor-doped insulating crystals.[33] There was a distinct lack of $Ir^{4+}$ related absorptions generally seen near the bandedge in insulating materials such as Zn- or Mg-doped β-$Ga_2O_3$,[44] indicating in these AGO samples the Fermi level may still be above mid-gap. There were no notable differences between VGF and CZ grown transmission measurements in 10% Al AGO.

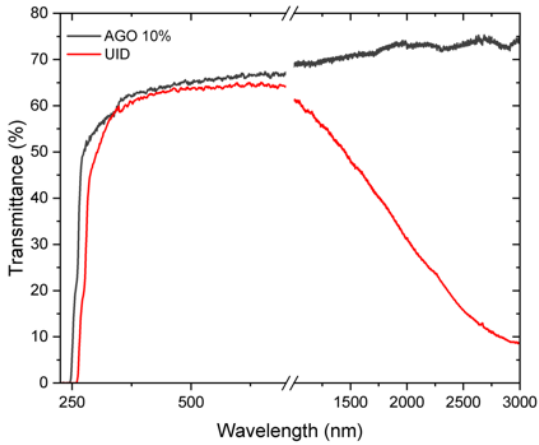

FIG 7. Transmission of 10% Al AGO and UID β-$Ga_2O_3$ samples, both approximately 1.5 mm thick, indicating lack of free carrier absorption in 10% Al AGO. An apparent absorption most notable in AGO 10% but also in UID at ≈350 nm was due to detector changeover, and is not a real feature.

Tauc analysis was used to quantitatively assess the shift in optical bandgap, $E_{opt}$, with the equation (1) used to calculate the absorption coefficient and the Tauc equation (2) for the bandgap.

$$\alpha = \frac{l}{L_1} \times \ln\left[\frac{(1-R)^2 \times 100}{2 T_1} + \sqrt{R^2 + \frac{(1-R)^4}{4(T_1/100)^2}}\right] \qquad (1)$$

$$(\alpha h\nu)^{\frac{1}{\gamma}} = B(h\nu - E_{opt}) \quad (2)$$

where the absorption coefficient α is determined from the transmission, the thickness, and calculated reflectance as determined by the refractive index (assumed to be identical to β-Ga$_2$O$_3$).[45] For the Tauc plot, $\gamma$ is a constant with a value of 0.5 for a direct bandgap, or 2 for an indirect bandgap. β-Ga$_2$O$_3$ is a material with a small (0.05 eV) difference between direct and indirect bandgap,[46,47] such that it can be treated as a direct bandgap,[47,48] thus 0.5 was selected as the value for $\gamma$. Analysis revealed a $\Delta E_{opt}$ +0.20 eV from UID to 10% Al AGO, which presented an $E_{opt}$ of ≈4.99 eV (Fig. 7). Polarization-dependent absorption as light is polarized parallel to the *b* or *c* planes was also observed; analysis of UID samples revealed an apparent bandgap of 4.51 eV and 4.79 eV for measurements polarization parallel to *c* and *b* respectively, and AGO demonstrates a bandgap of 4.79 eV and 4.99 eV for polarization parallel to *c* and *b* respectively. There were no appreciable differences in E$_{opt}$ between VGF and CZ 10% Al AGO samples. Comparison to theoretical work indicates such a $\Delta E_{opt}$ would occur with approximately 10 at.% Al content, verifying the alloying content present in the crystals.[5,49] Note that the measured apparent band gap for UID β-Ga$_2$O$_3$ is slightly low compared to theoretical, which is not surprising since even at ≈150 μm thick samples there is still some contribution from the Urbach tail absorption.

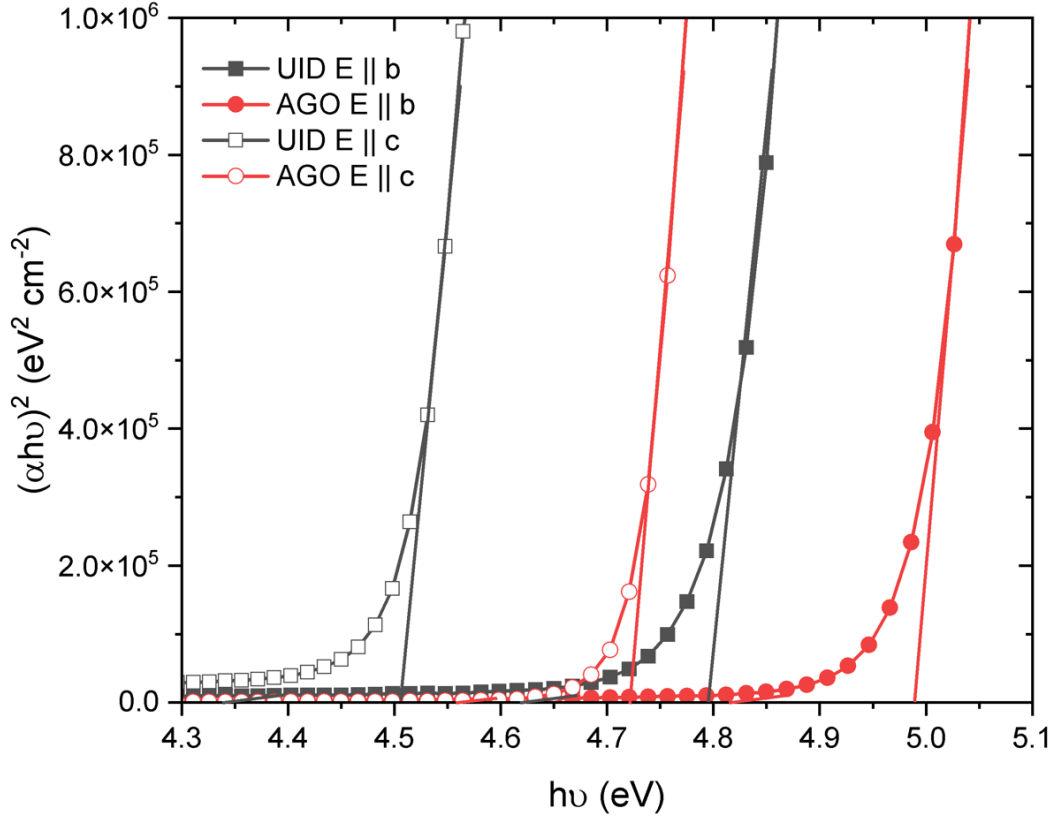

FIG. 8: Tauc plot of UID and 10% Al AGO, showing a $\Delta E_{opt}$ ~0.20 eV.

In order to compare band edge behavior between all AGO materials regardless of crystallinity, powder UV-Vis was also collected with an integrating sphere (Fig. 9). All three alloy concentrations show the optical band edge noticeably shifted from UID β-$Ga_2O_3$, shown in Fig. 9b. Powder UV-Vis transmittance data was transformed into Kubelka-Munk data, and Equation 3 and further analysis[50] was applied to study the bandgap,

$$(F(R_\infty) \times h\upsilon)^{\frac{1}{\gamma}} = B(h\upsilon - E_{opt}) \qquad (3)$$

where $F(R_\infty)$ is the Kubelka-Munk function, and the other variables are as described previously for equation (2). The $\Delta E_{opt}$ between UID and 10% Al AGO from Kubelka-Munk Tauc plot was +0.19 eV, whereas window Tauc plot was +0.20 eV. It is notable that bandgap bowing as a function of alumina concentration is seemingly observed in Figure 9b and 9c, which has been shown previously in theoretical work.[5] Bowing in this case was primarily of the conduction band edge and thus affects shallow donor behavior, since the valence band is still very flat and is not conducive to *p*-type behavior.[5] The bowing parameters $b$ were obtained from Equation 4, where $E_{opt}[Ga_2O_3]$ was the experimentally measured 4.52 eV and $E_{opt}[Al_2O_3]$ in either direct or indirect gap is obtained from Peelaers *et al.*[5] Bowing parameters calculated from Kubelka-Munk powder UV-Vis data were found to be 2.16 ± 0.16 eV if $Al_2O_3$ was direct and 1.69 ± 0.15 eV if $Al_2O_3$ was indirect. Calculations done by Peelaers *et al.* showed bowing parameters of 1.78 – 1.87 eV,[5] while thin film literature suggested broader results: 1.3,[51] 1.25[52] and 0.4 eV[53] have all been reported.

$$E_{opt}(x) = (1 - x)E_{opt}[Ga_2O_3] + xE_{opt}[Al_2O_3] - bx(1 - x) \qquad (4)$$

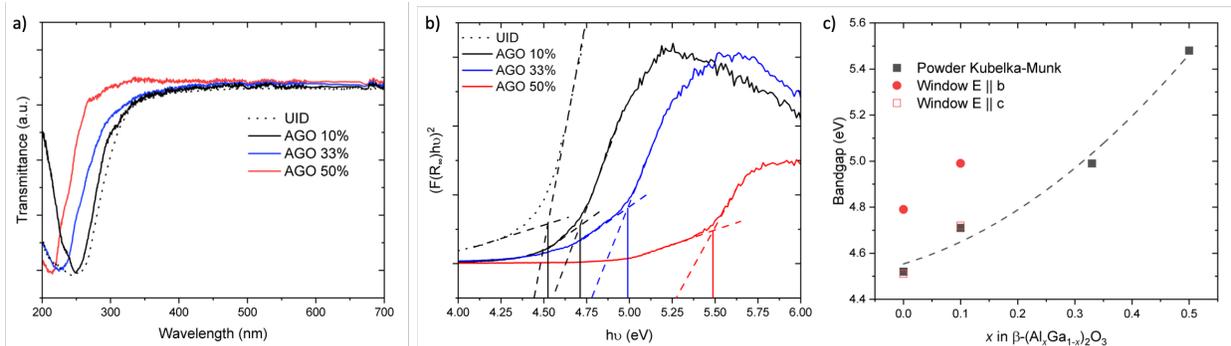

*FIG 9. a) Powder UV-Vis demonstrating a shift in the band edge as a function of alumina content. Transmittance has been artificially normalized to the band edge absorption peak for easier comparison. b) Kubelka-Munk Tauc plot powder samples shown in a). c) Kubelka-Munk powder UV-Vis and Window UV-Vis bandgap as a function of Al concentration. Line shown is a quadratic fit to Kubelka-Munk data.*

FTIR transmission measurements (Fig. 10) showed similar characteristics to β-Ga$_2$O$_3$ studied previously;[33] however, crystals showed a distinct lack of the ≈5155 cm$^{-1}$ Ir$^{4+}$ absorption peak related to 6-coordinated Ir$_{Ga}$,[44] which was found in material with a low Fermi level and thus lack of Ir$^{3+}$. Typically acceptor doped (e.g., Mg, Fe, and Zn) β-Ga$_2$O$_3$ demonstrated this Ir$^{4+}$ absorption peak.[33,44] 10% Al AGO showed a lack of free carrier concentration, which initially implied a low Fermi level, acceptor doping, and thus Ir$^{4+}$ formation; however, several samples were measured with FTIR across the VGF and CZ boule and all demonstrated a lack of the $Ir^{4+}$ peak. This indicated that the more insulating behavior of AGO (Fig. 10) was *not* due to an increase in acceptors deep in the gap – otherwise the $Ir^{4+}$ absorption would intensify, but instead the widening of the bandgap and deepening of shallow donor states near the conduction band edge was suggested.

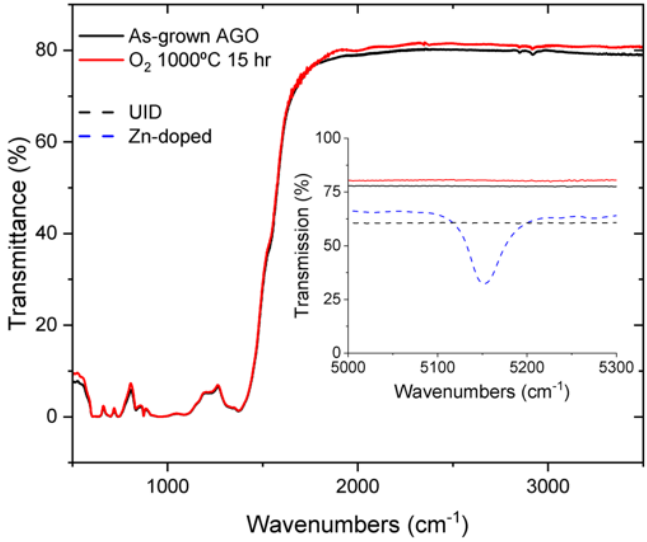

FIG 10. FTIR transmission on as-grown and annealed 10% Al AGO, inset showing Ir absorption region and UID, acceptor (Zn) doped, and AGO in that region.

### D. Electrical Properties

Resistivity of 50% Al AGO samples were not measured due to lack of single crystalline material. Only one sample of 33% Al AGO was measured, which reported $6\times10^7$ Ω·cm. 10% Al AGO showed insulating behavior, with an average resistivity higher than UID β-$Ga_2O_3$ (Fig. 11), which is also supported by a definitive lack of free carrier absorption in the near infrared (Fig. 7). These results were similar to those published by Li *et al.*, in which their optical floating zone grown 15% Al AGO was reported to have a resistivity of $1.5\times10^{12}$ Ω·cm.[29] This was most likely due to the raising of the conduction band minimum which converted the typical shallow donor impurities to deeper states,[12] equating to a lowering of the Fermi level. Further, Si concentration between UID crystals and AGO was similar, both being on the order of $10^{18}$ atoms $cm^{-3}$, and with no appreciably large acceptor concentration in AGO.

While Si was been shown by Varley[12] to remain an *n*-type shallow donor until much higher Al concentrations (>80% Al), it was been recently shown that shallow hydrogen donor states could become deep donors in AGO at Al concentrations as small as 1%.[54] Specifically, $H_i$ and $H_O$ could become deep centers, and compensate the Si donor impurity. The work by Mu *et al.* also discussed complexes of H with Si and C; however these were either unstable or were only acceptors at > 56% Al AGO.[54] Typically, the Si impurity in β-$Ga_2O_3$ materials was $10^{17} - 10^{18}$ atoms $cm^{-3}$, with Fe at similar concentrations; not much hydrogen would be needed to cause insulating behavior. Hydrogen may have seemed like a trace, or background impurity, but it was shown experimentally that in UID β-$Ga_2O_3$, hydrogen's contribution to free carrier concentration could be significant and was omnipresent in Czochralski grown crystals,[55] and thus the deepening of these shallow states could lead to insulating behavior.

The presence of hydrogen is ubiquitous due to the nature of Czochralski growth; the growth chamber is typically not well sealed especially to the small atom of hydrogen, and hydrogen can infiltrate into the source powder, insulation, or the chamber at many stages of growth preparation. An alternative explanation to the increased resistivity is a higher oxygen content in the melt environment due to inclusion of $Al_2O_3$, which is typically oxygen poor and supplemented by the gas flow scheme.[1,35] Such an oxygen rich environment would reduce the amount of oxygen related point defects, and potentially cause insulating behavior.

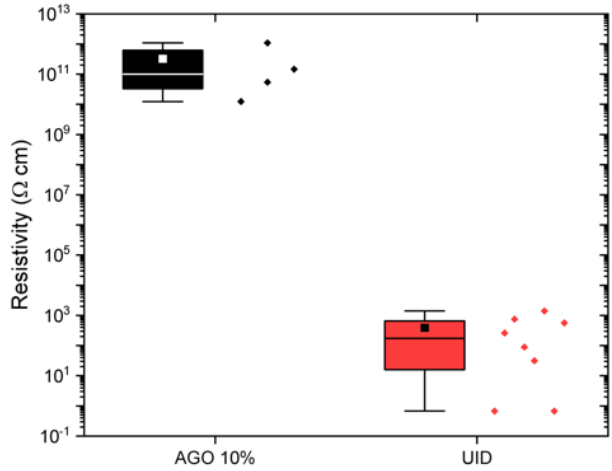

*FIG 11. Resistivity of 10% Al AGO and UID. Individual data points shown. For the statistics, the horizontal line in the box is the median, the small inner box is the mean, the colored box represents the 75%/25% interquartile range, and the short horizontal lines the 95%/5% range.*

## IV. Conclusion

Successful single crystal growth and incorporation of 10 mol.% $Al_2O_3$ in AGO (i.e., monoclinic β-$(Al_{0.1}Ga_{0.9})_2O_3$ or β-$Al_{0.2}Ga_{1.8}O_3$) has been demonstrated, as well as single phase polycrystalline growth of 33 AGO (i.e., monoclinic β-$(Al_{0.33}Ga_{0.67})_2O_3$ or β-$Al_{0.66}Ga_{1.34}O_3$) and 50 mol.% AGO (i.e., monoclinic β-$(Al_{0.5}Ga_{0.5})_2O_3$ or β-$AlGaO_3$). A β-$Ga_2O_3$+0.20 eV increase of the optical bandgap was demonstrated for β-$(Al_{0.1}Ga_{0.9})_2O_3$ from, to 4.99 eV. Crystal quality of 10% Al AGO is comparable to previously grown β-AGO, although somewhat lower quality than literature results for UID crystals, as shown by HAADF-STEM and high resolution rocking curve XRD, and quality could be improved for use as a substrate for thin film. However, we have demonstrated clear trends due to Al addition in β-AGO over a large range of $Al_2O_3$ (10-50 mol.%). By NMR we show that Al incorporated on both Ga sites, but favored the octahedral site ($Ga_{II}$), which becomes more disordered as Al content increases. HAADF-STEM further demonstrated Al incorporation onto both crystallographic sites, even for β-$(Al_{0.1}Ga_{0.9})_2O_3$. Disordering was also implied by the broadening of Raman vibrational bands at higher Al concentrations. The crystal lattice shrank linearly as a function of Al incorporation, as expected from the ionic size, and lattice empirical equations were generated allowing Al concentration to be estimated from these lattice parameters. Resistivity increased when alloying 10% $Al_2O_3$ with $Ga_2O_3$, and free carrier absorption in the near infrared was not observed, presumably due to a widening of the bandgap where normally shallow donors such as hydrogen become deeper states, or else related to a higher oxygen concentration in the melt when $Al_2O_3$ is introduced.[12,54] Overall, this study presented optical band gap, lattice parameters, aluminum site distribution, and vibrational spectra for a large range of single phase monoclinic β-AGO (to 50 mol.% $Al_2O_3$) which should be useful for future researchers.

## SUPPLEMENTARY MATERIAL

See the supplementary material for a larger XRD pattern figure, crystal purity data, as well as NMR fitting parameters and calculated site fractions.

## Acknowledgments

This material is based upon work supported by the Air Force Office of Scientific Research under award number FA9550-21-1-0078 monitored by Dr. Ali Sayir. Any opinions, finding, and conclusions or recommendations expressed in this material are those of the author and do not necessarily reflect the views of the United States Air Force. The authors thank David Bollinger for tireless discussion and advice concerning Rietveld refinement, as well as Brooke Downing for technical support. The authors also thank Billy Schmuck at the WSU for some assistance with XRD measurements. Finally, we also thank Arkka Bhattacharyya for discussions concerning AGO thin films and devices.

## Data Availability Statement

The data that support the findings of this study are available from the corresponding author upon reasonable request.

# REFERENCES

[1] Z. Galazka, R. Uecker, D. Klimm, K. Irmscher, M. Naumann, M. Pietsch, A. Kwasniewski, R. Bertram, S. Ganschow, and M. Bickermann, ECS Journal of Solid State Science and Technology **6**, Q3007 (2016).

[2] K. Hoshikawa, E. Ohba, T. Kobayashi, J. Yanagisawa, C. Miyagawa, and Y. Nakamura, Journal of Crystal Growth **447**, 36 (2016).

[3] A. Kuramata, K. Koshi, S. Watanabe, Y. Yamaoka, T. Masui, and S. Yamakoshi, Japanese Journal of Applied Physics **55**, 1202A2 (2016).

[4] Y. Tomm, J.M. Ko, A. Yoshikawa, and T. Fukuda, Solar Energy Materials and Solar Cells **66**, 369 (2001).

[5] H. Peelaers, J.B. Varley, J.S. Speck, and C.G. Van de Walle, Appl. Phys. Lett. **112**, 242101 (2018).

[6] T. Wang, W. Li, C. Ni, and A. Janotti, Phys. Rev. Applied **10**, 011003 (2018).

[7] T. Matsumoto, M. Aoki, A. Kinoshita, and T. Aono, Japanese Journal of Applied Physics **13**, 1578 (1974).

[8] W.S. Hwang, A. Verma, H. Peelaers, V. Protasenko, S. Rouvimov, H. (Grace) Xing, A. Seabaugh, W. Haensch, C.V. de Walle, Z. Galazka, M. Albrecht, R. Fornari, and D. Jena, Appl. Phys. Lett. **104**, 203111 (2014).

[9] M. Higashiwaki, K. Sasaki, A. Kuramata, T. Masui, and S. Yamakoshi, Appl. Phys. Lett. **100**, 013504 (2012).

[10] T. Minami, Y. Nishi, and T. Miyata, Applied Physics Express **6**, 044101 (2013).

[11] L.A.M. Lyle, S. Okur, V.S.N. Chava, M.L. Kelley, R.F. Davis, G.S. Tompa, M.V.S. Chandrashekhar, A.B. Greytak, and L.M. Porter, Journal of Electronic Materials **49**, 3490 (2020).

[12] J.B. Varley, A. Perron, V. Lordi, D. Wickramaratne, and J.L. Lyons, Appl. Phys. Lett. **116**, 172104 (2020).

[13] A. Hassa, H. von Wenckstern, L. Vines, and M. Grundmann, ECS Journal of Solid State Science and Technology **8**, Q3217 (2019).

[14] A.F.M. Anhar Uddin Bhuiyan, Z. Feng, J.M. Johnson, Z. Chen, H.-L. Huang, J. Hwang, and H. Zhao, Appl. Phys. Lett. **115**, 120602 (2019).

[15] T. Oshima, Y. Kato, N. Kawano, A. Kuramata, S. Yamakoshi, S. Fujita, T. Oishi, and M. Kasu, Applied Physics Express **10**, 035701 (2017).

[16] Y. Zhang, A. Neal, Z. Xia, C. Joishi, J.M. Johnson, Y. Zheng, S. Bajaj, M. Brenner, D. Dorsey, K. Chabak, G. Jessen, J. Hwang, S. Mou, J.P. Heremans, and S. Rajan, Appl. Phys. Lett. **112**, 173502 (2018).

[17] K. Ghosh and U. Singisetti, Journal of Materials Research **32**, 4142 (2017).

[18] S. Krishnamoorthy, Z. Xia, C. Joishi, Y. Zhang, J. McGlone, J. Johnson, M. Brenner, A.R. Arehart, J. Hwang, S. Lodha, and S. Rajan, Appl. Phys. Lett. **111**, 023502 (2017).

[19] C. Joishi, Y. Zhang, Z. Xia, W. Sun, A.R. Arehart, S. Ringel, S. Lodha, and S. Rajan, IEEE Electron Device Letters **40**, 1241 (2019).

[20] P. Ranga, A. Bhattacharyya, A. Chmielewski, S. Roy, R. Sun, M.A. Scarpulla, N. Alem, and S. Krishnamoorthy, Applied Physics Express **14**, 025501 (2021).

[21] P. Ranga, A. Bhattacharyya, A. Rishinaramangalam, Y.K. Ooi, M.A. Scarpulla, D. Feezell, and S. Krishnamoorthy, Applied Physics Express **13**, 045501 (2020).

[22] P. Ranga, A. Rishinaramangalam, J. Varley, A. Bhattacharyya, D. Feezell, and S. Krishnamoorthy, Applied Physics Express **12**, 111004 (2019).

[23] T. Oshima, T. Okuno, N. Arai, Y. Kobayashi, and S. Fujita, Japanese Journal of Applied Physics **48**, 070202 (2009).

[24] S.W. Kaun, F. Wu, and J.S. Speck, Journal of Vacuum Science & Technology A **33**, 041508 (2015).

[25] A.F.M. Anhar Uddin Bhuiyan, Z. Feng, J.M. Johnson, H.-L. Huang, J. Hwang, and H. Zhao, Crystal Growth & Design **20**, 6722 (2020).

[26] A.F.M.A.U. Bhuiyan, Z. Feng, J.M. Johnson, H.-L. Huang, J. Hwang, and H. Zhao, Appl. Phys. Lett. **117**, 142107 (2020).

[27] Z. Galazka, S. Ganschow, A. Fiedler, R. Bertram, D. Klimm, K. Irmscher, R. Schewski, M. Pietsch, M. Albrecht, and M. Bickermann, Journal of Crystal Growth **486**, 82 (2018).

[28] Z. Galazka, R. Schewski, K. Irmscher, W. Drozdowski, M.E. Witkowski, M. Makowski, A.J. Wojtowicz, I.M. Hanke, M. Pietsch, T. Schulz, D. Klimm, S. Ganschow, A. Dittmar, A. Fiedler, T. Schroeder, and M. Bickermann, Journal of Alloys and Compounds **818**, 152842 (2020).

[29] Z. Li, J. Chen, H. Tang, Z. Zhu, M. Gu, J. Xu, L. Chen, X. Ouyang, and B. Liu, ACS Appl. Electron. Mater. **3**, 4630 (2021).

[30] D.A. Bauman, D.Iu. Panov, D.A. Zakgeim, V.A. Spiridonov, A.V. Kremleva, A.A. Petrenko, P.N. Brunkov, N.D. Prasolov, A.V. Nashchekin, A.M. Smirnov, M.A. Odnoblyudov, V.E. Bougrov, and A.E. Romanov, Physica Status Solidi (a) **218**, 2100335 (2021).

[31] D. Zakgeim, D. Bauman, D. Panov, V. Spiridonov, A. Kremleva, A. Smirnov, M. Odnoblyudov, A. Romanov, and V. Bougrov, Applied Physics Express **15**, 025501 (2022).

[32] M. Saleh, J.B. Varley, J. Jesenovec, A. Bhattacharyya, S. Krishnamoorthy, S. Swain, and K. Lynn, Semiconductor Science and Technology **35**, 04LT01 (2020).

[33] J. Jesenovec, J. Varley, S.E. Karcher, and J.S. McCloy, Journal of Applied Physics **129**, 225702 (2021).

[34] M. Saleh, A. Bhattacharyya, J.B. Varley, S. Swain, J. Jesenovec, S. Krishnamoorthy, and K. Lynn, Applied Physics Express **12**, 085502 (2019).

[35] Z. Galazka, Journal of Applied Physics **131**, 031103 (2022).

[36] D. Massiot, F. Fayon, M. Capron, I. King, S. Le Calvé, B. Alonso, J.-O. Durand, B. Bujoli, Z. Gan, and G. Hoatson, Magnetic Resonance in Chemistry **40**, 70 (2002).

[37] V.G. Hill, R. Roy, and E.F. Osborn, Journal of the American Ceramic Society **35**, 135 (1952).

[38] K. Hoshikawa, T. Kobayashi, E. Ohba, and T. Kobayashi, Journal of Crystal Growth **546**, 125778 (2020).

[39] D.S. Cook, J.E. Hooper, D.M. Dawson, J.M. Fisher, D. Thompsett, S.E. Ashbrook, and R.I. Walton, Inorg. Chem. **59**, 3805 (2020).

[40] J.M. Johnson, H.-L. Huang, M. Wang, S. Mu, J.B. Varley, A.F.M.A. Uddin Bhuiyan, Z. Feng, N.K. Kalarickal, S. Rajan, H. Zhao, C.G. Van de Walle, and J. Hwang, APL Materials **9**, 051103 (2021).

[41] T. Onuma, S. Fujioka, T. Yamaguchi, Y. Itoh, M. Higashiwaki, K. Sasaki, T. Masui, and T. Honda, Journal of Crystal Growth **401**, 330 (2014).

[42] B. Lafuente, R.T. Downs, H. Yang, N. Stone, T. Armbruster, and R.M. Danisi, in (W. De Gruyter, Berlin, Germany, 2015), pp. 1–30.

[43] J.L. Lyons, Semiconductor Science and Technology **33**, 05LT02 (2018).

[44] J.R. Ritter, K.G. Lynn, and M.D. McCluskey, Journal of Applied Physics **126**, 225705 (2019).

[45] I. Bhaumik, R. Bhatt, S. Ganesamoorthy, A. Saxena, A.K. Karnal, P.K. Gupta, A.K. Sinha, and S.K. Deb, Appl. Opt. **50**, 6006 (2011).

[46] C. Janowitz, V. Scherer, M. Mohamed, A. Krapf, H. Dwelk, R. Manzke, Z. Galazka, R. Uecker, K. Irmscher, R. Fornari, M. Michling, D. Schmeißer, J.R. Weber, J.B. Varley, and C.G.V. de Walle, New Journal of Physics **13**, 085014 (2011).

[47] H. Peelaers and C.G. Van de Walle, Physica Status Solidi (b) **252**, 828 (2015).

[48] K.A. Mengle, G. Shi, D. Bayerl, and E. Kioupakis, Appl. Phys. Lett. **109**, 212104 (2016).

[49] J.B. Varley, Journal of Materials Research (2021).

[50] P. Makuła, M. Pacia, and W. Macyk, J. Phys. Chem. Lett. **9**, 6814 (2018).

[51] J. Li, X. Chen, T. Ma, X. Cui, F.-F. Ren, S. Gu, R. Zhang, Y. Zheng, S.P. Ringer, L. Fu, H.H. Tan, C. Jagadish, and J. Ye, Appl. Phys. Lett. **113**, 041901 (2018).

[52] A.F.M.A.U. Bhuiyan, Z. Feng, J.M. Johnson, H.-L. Huang, J. Hwang, and H. Zhao, Appl. Phys. Lett. **117**, 252105 (2020).

[53] J. Bhattacharjee, S. Ghosh, P. Pokhriyal, R. Gangwar, R. Dutt, A. Sagdeo, P. Tiwari, and S.D. Singh, AIP Advances **11**, 075025 (2021).

[54] S. Mu, M. Wang, J.B. Varley, J.L. Lyons, D. Wickramaratne, and C.G.V. de Walle, (2021).

[55] Z. Galazka, K. Irmscher, R. Schewski, I.M. Hanke, M. Pietsch, S. Ganschow, D. Klimm, A. Dittmar, A. Fiedler, T. Schroeder, and M. Bickermann, Journal of Crystal Growth **529**, 125297 (2020).